\documentstyle[12pt]{article}
\begin{document}
\setlength{\baselineskip}{0.9 cm}
\hfill May, 1997
\vskip 0.5in
\centerline{\bf \large Gaussian Time-Dependent Variational Principle for Bosons}
\centerline{\bf{I - Uniform Case}}
\vskip 0.25in
\centerline{Arthur K. Kerman}
\centerline{Center for Theoretical Physics, Laboratory for
Nuclear Science}
\centerline{       and Department of Physics}
\centerline{   Massachusetts Institute of Technology}
\centerline{   Cambridge, MA 02139}
\vskip 0.1in
\centerline{Paolo Tommasini}
\centerline{Institute for Theoretical Atomic and Molecular Physics}
\centerline{Harvard-Smithsonian Center for Astrophysics}
\centerline{Cambridge, MA 02138}
\vskip 0.15in
\centerline{                ABSTRACT}
\bigskip

We investigate the Dirac time-dependent variational method for a system of
non-ideal Bosons interacting through an arbitrary two body potential.
The method produces a set of non-linear time dependent
equations for the variational parameters. In particular
 we have considered small
oscillations about equilibrium.
We obtain  generalized RPA equations that can be understood as interacting 
quasi-bosons, usually mentioned in the
literature as having a gap. The result of this interaction provides us 
with scattering properties of these quasi-bosons
including  possible bound-states, which can include zero modes. In fact the 
zero mode bound state  can
be interpreted as a new quasi-boson with a gapless dispersion relation.
Utilizing these results we discuss  a straightforward scheme for introducing temperature.
\section{Introduction}

 The objective of this paper is to exhibit the most general way of obtaining
time dependent equations of motion in the Gaussian approximation \cite{KK}. This 
will lead to the so called generalized RPA, when one 
examines infinitesimal oscillations about equilibrium.
The static solution in the uniform case
 can be obtained using several other methods
\cite{GG} - \cite{HP} leading to a gap in the quasi-boson energy.
We show here that  the time dependent RPA equations lead to a gapless mode.
In
fact this {\bf must}  happen because  particle number conservation symmetry was broken 
in the static solution, so the zero gap
 is exactly the
  associated Goldstone mode. This can be seen as an alternative to
the
functional derivative \cite{HM} method in the Girardeau-Arnowitt \cite{GR}
approximation. We stress that to obtain this result the equations of motion
must arise from a consistent variational scheme. An important element of this
scheme is the fact that even in the uniform case the variational quantities
have to depend on two momenta. It is actually this feature which leads to a
 gapless result. Since this solves the problem of the gap, truncations normally
 used to avoid it \cite{TR} are not necessary.

We  would like to point out that due to the likelihood of increased densities in
future  experiments on Bose-Einstein condensation in atomic traps
\cite{AE} - \cite{KM1} our time dependent equations may  be  useful
 to study non-equilibrium evolution.
 The RPA equations also give us the possibility of investigating  the scattering
of two quasi-bosons in the medium.
Information about  scattering  can be used in calculating the free energy at finite 
temperature \cite{RD} which  takes into account the gapless dispersion relation in contrast to the so called
Temperature Hartree Fock Bogoliubov method \cite{TT}.

The structure of this paper is as follows. Section 2 reviews the time-dependent
 variational principle and the canonical nature of the equations of motion
arising from it. In section 3 we discuss the static solution. Section 4
obtains the small oscillation approximation (RPA) for the equations of motion. The
Goldstone mode is discussed in Section 5. In section 6 we show the  connection with the Bogoliubov transformation 
and calculate the one and two quasi-boson  wave functions. In section 7 we show how the two quasi-boson scattering affects the thermodynamics  using well know methods os statistical mechanics.

\section{General Formalism}
In this section we shall review \footnote{This material is substantially the same as some sections of \cite{KP} and it is included here for completeness}
 Time-dependent
Variational Principle {\cite{KK}}-{\cite{KL}} and show how it can be 
implemented in the
non-relativistic many-body case. First we define an effective action 
functional for the
time-dependent quantum system
\begin{equation}
\label{1}
S = \int L(t) dt = \int dt \langle \Psi,t | (i \partial_{t}
- \hat{H}) | \Psi, t \rangle,
\end{equation}
\noindent where $| \Psi,t \rangle$ is the quantum state of 
the system and
$\hat{H}$ is the Hamiltonian of the theory. For a system of
non-relativistic
interacting Bosons we have [we use the notation : $\int_{\bf
 x} = \int d^{3}x$]
\begin{equation}
\label{2}
\hat{H} = \int_{{\bf x},{\bf y}}  \hat{\psi}({\bf x})^{\dag}
 h({\bf x},{\bf y}) \hat{\psi}({\bf y})  + \frac{1}{2}\int_{
{\bf x},{\bf y}}
\hat{\psi}({\bf y})^{\dag}\hat{\psi}({\bf x})^{\dag} V({\bf
x} - {\bf y})
\hat{\psi}({\bf x}) \hat{\psi}({\bf y}),
\end{equation}
\noindent where the one body Hamiltonian $h({\bf x},{\bf y})
$ may include a one
 body external potential. The operators $\hat{\psi}$ and 
$\hat{\psi}^{\dag}$
can be written in the form
\begin{eqnarray}
\label{3}
\hat{\psi}({\bf x}) &=& \frac{1}{\sqrt{2}} \left[ \hat{\phi}
({\bf x}) + i \hat{\pi}({\bf x}) \right] \\
\label{4}
\hat{\psi}({\bf x})^{\dag} &=& \frac{1}{\sqrt{2}} \left[ 
\hat{\phi}({\bf x}) -i \hat{\pi}({\bf x}) \right]
\end{eqnarray}
\noindent where $\hat{\phi}({\bf x})$ is the field operator
and
$\hat{\pi}({\bf x})$ the canonical field momemtum

We can obtain the time dependent Schr\"odinger equation by 
requiring that $S$
is stationary, supplemented by appropriate boundary 
conditions,
 under the most general variation of $| \Psi ,t \rangle$.
The variational scheme is implemented by chosing  a trial wave functional
describing the
system. Working in the functional Shr\"odinger picture we 
replace the abstract
state $| \Psi ,t \rangle$ by a wave functional of the field
$\phi'({\bf x})$
\begin{equation}
\label{5}
| \Psi,t \rangle \rightarrow \Psi [ \phi',t].
\end{equation}
\noindent The action of the operators $\hat{\phi}({\bf x})$
and the canonical momentum
$\hat{\pi}({\bf x})$ are realized respectively by
\begin{eqnarray}
\label{6}
\hat{\phi}({\bf x}) |\Psi,t \rangle & \rightarrow & \phi'({
\bf x}) \Psi[\phi',t]  \\
\label{7}
\hat{\pi}({\bf x}) | \Psi,t \rangle & \rightarrow & - i 
\frac{\delta}{\delta
\phi'({\bf x})} \Psi[\phi',t].
\end{eqnarray}
\noindent The mean value of any operator is calculated by 
the functional integral
\begin{equation}
\label{8}
\langle \Psi,t |{ \cal O} | \Psi , t \rangle = \int ({\cal D
} \phi') \Psi^{\ast}
[\phi' , t] {\cal O} \Psi[\phi' , t],
\end{equation}
\noindent where $\Psi$ is normalized to unity. The Gaussian
approximation
consists in taking a Gaussian trial wave functional in its 
most general
parametrization
\begin{eqnarray}
\label{9}
\Psi[\phi',t] &=& N \exp\left\{ - \int_{{\bf x},{\bf y}}  
\delta \phi'({\bf x},t) \left[\frac{G^{-1}({\bf x},{\bf y},t)}
{4} - i \Sigma({\bf x},{\bf y},t) \right] \delta \phi'({\bf
y},t) \right. \nonumber \\
&& + \left.i \int_{\bf x}  \pi({\bf x},t)\delta\phi'({\bf x}
,t)
 \right\},
\end{eqnarray}
\noindent
with $\delta \phi'({\bf x},t) = \phi'({\bf x}) - \phi({\bf x
},t) $. Due to the fact that the Hamiltonian commutes with 
the number of particles
\begin{equation}
\label{10}
[\hat{H},\hat{N}] =0
\end{equation}
\noindent we can define a more general trial functional
\begin{equation}
\label{11}
| \Psi',t \rangle  = e^{-i \hat{N} \theta(t)}  |\Psi,t 
\rangle,
\end{equation}
\noindent where $\theta(t)$ is another variational parameter
 introduced because of this continuous symmetry.
Thus  our variational parameters are $\phi({\bf x},t)$,
$\pi({\bf x},t)$, $\theta(t)$,$G({\bf x},{\bf y},t)$ and
$\Sigma({\bf x},{\bf y},t)$ ,with  $G$ and $\Sigma$ being  
real symmetric matrices.
These quantities are related to the following mean-values:
\begin{eqnarray}
\label{12}
\langle \Psi' , t | \hat{\phi}({\bf x}) | \Psi' , t 
\rangle
 & = & \phi({\bf x},t) \\
\label{13}
\langle \Psi' , t | \hat{\pi}({\bf x}) | \Psi' , t \rangle
 & = & \pi({
\bf x},t) \\
\label{14}
\langle \Psi' , t | \hat{\phi}({\bf x})  \hat{\phi}({\bf y
}) | \Psi' , t \rangle & = & G({\bf x},{\bf y},t) + \phi({
\bf x},t) \phi({\bf y},t) \\
\langle \Psi' , t | \hat{\pi}({\bf x},t)  \hat{\pi}({\bf y
},t) | \Psi' , t \rangle & = & \frac{G^{-1}({\bf x},{\bf y},t)}{4}
+4 \int_{{\bf w},{\bf z}}  \Sigma({\bf x},{\bf w},t) G({
\bf w},{\bf z},t) \Sigma({\bf z},{\bf y},t) \nonumber  \\
&&+ \pi({\bf x},t) \pi({\bf y},t) \\
\label{15a}
i \langle \Psi' , t | \hat{\phi}({\bf x},t)  \hat{\pi}({\bf y
},t) | \Psi' , t \rangle & = & -\frac{\delta({\bf x}-{\bf y})}{2} + i \int_{\bf z} \left[\Sigma({\bf x},{\bf z},t) G({\bf z},{\bf y},t) + G({\bf y},{\bf z},t) \Sigma({\bf z},{\bf x},t) \right] \nonumber \\
&&+ i \phi({\bf x},t) \pi({\bf y},t)  \\
\label{15}
\langle \Psi' , t |i \frac{\delta}{\delta t}  | \Psi' , t 
\rangle  & = &
\int_{{\bf x},{\bf y}}  \Sigma({\bf x},{\bf y},t) \dot{G}({
\bf y},{\bf x},t) + \int_{\bf x}  \pi({\bf x},t) \dot{\phi}({
\bf x},t)  \nonumber \\
&&+{\cal N} \dot{\theta}(t) + \mbox{ total time derivatives}.
\end{eqnarray}
\noindent We may
 ignore the total time derivatives because they do not
contribute to the equations of motion. If now we write the 
action we will get
\begin{equation}
\label{16}
S = \int dt \left(\int_{\bf x} \pi({\bf x},t)\dot{\phi}({\bf
 x},t) +
\int_{{\bf x},{\bf y}} \Sigma({\bf x},{\bf y},t) \dot{G}({
\bf y},{\bf x},t) +
{\cal N} \dot{\theta} (t) - {\cal H} \right),
\end{equation}
\noindent where
\begin{equation}
\label{17}
{
\cal H} =\langle \Psi' , t | \hat{H}| \Psi' , t \rangle
\end{equation}
\noindent and
\begin{equation}
{\cal N} = \langle \Psi' , t | \hat{N}| \Psi' , t \rangle.
\end{equation}
\noindent From (\ref{16}) we see that $({\cal N},\theta)$, $
(\pi,\phi)$ and
$(\Sigma,G)$ are canonical pairs.
Because of the symmetry ${\cal H}$ has no dependence on $
\theta$ and it
follows that
$\dot{N} =0$ and $\dot{\theta}(t) =$ constant $ \equiv \mu $
. We can now write
the remaining Hamilton equations,
\begin{eqnarray}
\label{18}
\dot{\phi}({\bf x},t) &=& \frac{\delta( {\cal H} - \mu {\cal
 N})}{\delta \pi ({\bf x},t)},\\
\label{19}
\dot{\pi}({\bf x},t) &=& - \frac{\delta( {\cal H} - \mu {
\cal N})}{\delta \phi ({\bf x},t)},\\
\label{20}
\dot{G}({\bf x},{\bf y},t) &=& \frac{\delta( {\cal H} - \mu
{\cal N})}{\delta
\Sigma({\bf x},{\bf y},t)}, \\
\label{21}
\dot{\Sigma}({\bf x},{\bf y},t) &=&- \frac{\delta( {\cal H}
- \mu {\cal N})}{\delta
G({\bf x},{\bf y},t)}.
\end{eqnarray}
\noindent For convenience we introduce
\begin{equation}
\label{22}
\psi({\bf x},t) \equiv \langle \hat{\psi}({\bf x}) \rangle =
 \frac{\phi({\bf x},t) + i \pi({\bf x},t)}{\sqrt{2}},
\end{equation}
\noindent so that Eqs.(\ref{18})-(\ref{19})  become
\begin{equation}
\label{23}
\imath \dot{\psi}({\bf x},t) = \frac{\delta( {\cal H} - 
\mu {\cal N})}{\delta \psi^{\ast}({\bf x},t)}
\end{equation}
\noindent It is well known that a Gaussian 
functional leads to the mean field factorization \cite{GG} i.e. 
\begin{eqnarray}
\label{26}
{\cal H} - \mu {\cal N} = &\int_{{\bf x},{\bf y}}& \left\{\left[h({
\bf x},{\bf y}) - \mu \delta({\bf x}-{\bf y})  \right]  \rho({\bf x},{\bf y},t)
+ \frac{1}{2} V({\bf x}-{\bf y})
|\psi({\bf x},t)|^{2}  |\psi({\bf y},t)|^{2}
 \right\}\nonumber \\
+ \frac{1}{2} & \int_{{\bf x},{\bf y}}& V({\bf x}-{\bf y}) \left[
R({\bf y},{\bf x},t) R({\bf x},{\bf y},t) +
R({\bf x},{\bf x},t) R({\bf y },{\bf y},t) +
D^{\ast} ({\bf x},{\bf y},t) D({\bf x},{\bf y},t) \right] \nonumber \\
+  &\int_{{\bf x},{\bf y}}& V({\bf x}-{\bf y}) \left[
  \frac{1}{2} \psi^{\ast}({\bf x},t)\psi({\bf y},t) R({\bf x},{\bf y},t)
+  \frac{1}{2} \psi^{\ast}({\bf y},t)\psi({\bf x},t) R({\bf y },{\bf x},t)
+  |\psi({\bf x},t)|^{2} R({\bf y },{\bf y},t) \right] \nonumber \\
- \frac{1}{2} &\int_{{\bf x},{\bf y}}&  V({\bf x}-{\bf y}) \left[
\psi({\bf x},t) \psi({\bf y},t) D^{\ast}({\bf x},{\bf y},t) +
\psi^{\ast}({\bf x},t) \psi^{\ast}({\bf y},t) D({\bf x},{\bf y},t)
\right]
\end{eqnarray}
\noindent and 
\begin{eqnarray}
\label{27}
\rho({\bf x},{\bf y},t) &=& \langle \psi^{\dag}({\bf x}) 
\psi({\bf y}) \rangle= \psi^{\ast}({\bf x},t) \psi({\bf y},t)
+ R({\bf x},{\bf y},t)\\
\label{28}
\Delta({\bf x},{\bf y},t) &=&-\langle \psi({\bf x}) \psi({\bf
 y}) \rangle = - \psi({\bf x},t) \psi({\bf y},t) + D({\bf x}
,{\bf y},t)
\end{eqnarray}
\noindent with
\begin{eqnarray}
\label{29}
R({\bf x},{\bf y},t) &=& \frac{1}{2} \left[ \frac{G^{-1}({
\bf x},{\bf y},t)}
{4} + G({\bf x},{\bf y},t) - \delta({\bf x}-{\bf y})
\right]  
+2 \int_{{\bf w},{\bf z}}  \Sigma({\bf x},{\bf w},t) G({
\bf w},{\bf z},t)
\Sigma({\bf z},{\bf y},t)  \nonumber \\
&& + i \int_{\bf z} \left[G({\bf x},{\bf z},t) \Sigma({\bf z},{\bf y},t) - 
\Sigma({\bf x},{\bf z},t) G({\bf z},{\bf y},t) \right] \\
\label{30}
D({\bf x},{\bf y},t) &=& \frac{1}{2} \left[ \frac{G^{-1}({
\bf x},{\bf y},t)}{4} -G({\bf x},{\bf y},t) \right]
+  2 \int_{{\bf w},{\bf z}}
 \Sigma({\bf x},{\bf w},t) G({\bf w},{\bf z},t) \Sigma({\bf
z},{\bf y},t) \nonumber   \\
&& -i \int_{{\bf z}} \left[\Sigma({\bf x},{\bf z},t) G({\bf z},{\bf y},t) +
 G({\bf x},{\bf z},t)\Sigma({\bf z},{\bf y},t)\right],
\end{eqnarray}
\noindent because of (\ref{12})-(\ref{14}). We note that the
 density gets contributions from the condensate field
$\psi$ as well as from the fluctuations ($G$,$\Sigma$). The
contribution from
$\psi^{\ast} \psi$ is the condensate density.

We introduce the generalized potentials
\begin{eqnarray}
\label{32}
{\cal U}_{{\rm d}}({\bf x},{\bf y},t) &=& \delta({\bf x}-{
\bf y}) \int_{\bf z} \rho({\bf z},{\bf z},t) V({\bf x} -{\bf z
}) \\
\label{31}
{\cal U}_{{\rm e}}({\bf x},{\bf y},t) & = & \rho ({\bf x},{
\bf y},t) V({\bf x}-{
\bf y}) \equiv {\cal U}_{{\rm e}}^{\rm r } +i\; {\cal U}_{{\rm e}}^{\rm i} \\
\label{33}
{\cal U}_{{\rm p}} ({\bf x},{\bf y},t) &=& \Delta({\bf x},{
\bf y},t) V({\bf x}-{\bf y}) \equiv {\cal U}_{{\rm p}}^{\rm r
} +i\; {\cal U}_{{\rm p}}^{\rm i}
\end{eqnarray}
\noindent where the notation emphasizes real and imaginary parts of ${\cal U}_{\rm p}$. We also define the matrices
\begin{eqnarray}
\label{42}
C({\bf x},{\bf y},t) &=& h({\bf x},{\bf y}) -\mu + {\cal U}_
{e}({\bf x},{\bf y},
t) +{\cal U}_{d}({\bf x},{\bf y},t) \\
\label{40}
A({\bf x},{\bf y},t) &=& C^{\rm r}({\bf x},{\bf y},t) + {\cal U}_{\rm p}^{\rm r}({\bf x},{\bf y},t) \\
\label{41}
B({\bf x},{\bf y},t) &=& C^{\rm r}({\bf x},{\bf y},t)  - {\cal U}_{\rm p}^{\rm r}({\bf x},{\bf y},t) 
\end{eqnarray}
\noindent From Eq. (\ref{23})-(\ref{42}) we obtain 
 an abstract matrix form of the equations of motion
\begin{eqnarray}
\label{44}
\dot{\Sigma} &=& \frac{1}{8} G^{-1} A G^{-1} - 2 \Sigma A 
\Sigma - \frac{B}{2} +\{{\cal U}_{{\rm p}}^{\rm i},\Sigma\} - [{\cal U}_{\rm e}^{\rm i},\Sigma ] \\
\label{45}
\dot{G} &=&  \{A,\{G,\Sigma\}\}  -  \{{\cal U}_{{\rm p}}^{
\rm i},G \} - [[{\cal U}_{\rm e}^{\rm i},G] \\
\label{46}
\imath \dot{\psi} &=& C \psi -  {\cal U}_{{\rm p}}  \psi^{
\ast},
\end{eqnarray}
\noindent where
 we have used the fact that ($\Sigma$,G) are symmetric 
matrices. 
\noindent These equations (\ref{44})-(\ref{46})  are the nonlinear 
field equations for an
arbitrary interaction $V$ between the particles and
contain any external potential through  $h$. As an example 
the matrix product   $G^{-1} A G^{-1}$ can be written in 
coordinate representation as
\begin{equation}
\int_{{\bf z},{\bf w}}  G^{-1} ({\bf x},{\bf z},t) A({\bf z}
,{\bf w}) G^{-1} ({\bf w},{\bf y},t).
\end{equation}
The static equations  can be obtained by setting the canonical momenta to 
zero, that is $\Sigma({\bf x},{\bf t},t) = \pi({\bf x},t) =0$,
 $\dot{G}({\bf x},{\bf y},t) ={\dot{ \phi}}({\bf x},
t)=0$. From (\ref{22}), (\ref{29}) and (\ref{30}) we than 
have
\begin{eqnarray}
\label{47}
R({\bf x},{\bf y},0)& \equiv & R({\bf x},{\bf y}) = \frac{1}
{2} \left[ \frac{G^{-1}({\bf x},{\bf y})}{
4} + G({\bf x},{\bf y}) - \delta({\bf x}-{\bf y}) \right] \\
\label{48}
D({\bf x},{\bf y},0) & \equiv & D({\bf x},{\bf y}) =\frac{1}
{2} \left[ \frac{G^{-1}({\bf x},{\bf y})}{4} -G({\bf x},{\bf
 y})\right] \\
\label{48i}
\psi({\bf x},0) & \equiv & \psi({\bf x}) =\frac{\phi({\bf x}
)}{\sqrt{2}}.
\end{eqnarray}
\noindent
So that for the static case we have to self consistently solve
\begin{eqnarray}
\label{49}
&&\frac{1}{4} \int_{{\bf z},{\bf w}} G^{-1}({\bf x},{\bf z})
 A({\bf z},{\bf w}) G^{-1}({\bf w},{\bf y}) - B({\bf x},{\bf
 y}) = 0 \\
\label{50}
&& \int_{\bf z} \left[ B({\bf x},{\bf z}) \psi({\bf z}) - 2
\psi({\bf x}) \psi^{2}({\bf z}) V({\bf x}-{\bf z}) \right] =
 0.
\end{eqnarray}
\noindent using (\ref{32})-(\ref{42}). We note that the constraint $G=1/2$ for the static solution leads to $R=D=0$ and $A=B$ so that equation (\ref{50}) becomes the usual non linear equation for the single quantity $\psi$ obtained from a many body permanent for bosons \cite{PT}. However the time dependent eqs. (\ref{44})-(\ref{46}) are more general, because our trial Gaussian is actually a coherent state with an indefinite number of particles. 

\section{Uniform Static Solution} 

To solve  equations (\ref{49}) and (\ref{50}) in the uniform case we  go to the momentum basis  where $G$ is 
diagonal simultaneously with $A$, $B$ and $C$. There exists a straightforward generalization for the nonuniform case where $A$, $B$ and $C$ do not commute.  
 In the uniform case  
$\phi({\bf k}) = \phi \delta({\bf k})$. The static equations (\ref{49}) and (\ref{50}) become
\begin{eqnarray} 
\label{51}
\frac{1}{4} G^{-2}({\bf k}) A({\bf k}) -B({\bf k})&=&0  \\ 
\label{52}
\phi \left[B(0) - \lambda \phi^{2} \right] &=&0.
\end{eqnarray}
\noindent Where we have defined  $\lambda \equiv \widetilde{V}(0) $, $\tilde{V}$ indicates the Fourier transform of the potential and  $e({\bf k}) \equiv \frac{\hbar^{2} k^2}{2 m}$. The diagonal matrices $A$ and $B$  (\ref{40})-(\ref{41}) can be written as 
\begin{eqnarray}
\label{53}
A({\bf k}) &=& e({\bf k}) - \mu + {\cal U}_{{\rm p}}^{\rm r} ({\bf k}) + {\cal U}_{{\rm e}}({\bf k}) + {\cal U}_{{\rm d}}({\bf k})  \\
\label{54}
B({\bf k}) &=& e({\bf k}) - \mu - {\cal U}_{{\rm p}}^{\rm r} ({\bf k}) + {\cal U}_{{\rm e}}({\bf k}) + {\cal U}_{{\rm d}}({\bf k}).
\end{eqnarray}
\noindent In this representation the generalized potentials (\ref{31})-(\ref{33}) become [we use the notation : $\int_{{\bf k}} = (2 \pi)^{-3} \int d^{3} k$ ] 
\begin{eqnarray}
\label{56}
{\cal U}_{\rm d}({\bf k}) &=& \lambda  \int_{{\bf k}'} \rho({\bf k}') \\
\label{55}
{\cal U}_{\rm e}({\bf k}) &=& \int_{{\bf k}'} \rho({\bf k}') \widetilde{V}({\bf k} - {\bf k}') \\
\label{57}
{\cal U}_{\rm p}({\bf k}) &=& \int_{{\bf k}'} \Delta({\bf k}') \widetilde{V}({\bf k} -
 {\bf k}') 
\end{eqnarray} 
\noindent The solutions for (\ref{51}) and (\ref{52}) are 
\begin{eqnarray}
\label{60}
&&G({\bf k})= \frac{1}{2} \sqrt{\frac{A({\bf k})}{B({\bf k})}}  \\ 
\label{61}
&&\phi=0 \;\;\;\;  
{\rm \underline{or} } \;\;\; B(0) =  \lambda \phi^{2}, 
\end{eqnarray}
\noindent where the second solution is symmetry breaking because $\langle  \psi \rangle = \phi/\sqrt{2} \ne 0$.
For stability reasons we shall see (in section 4) that both $A$ and $B$ must be positive which means that using (\ref{61}) there is no $\phi \ne  0$ solution for $\lambda  <  0$.
Using (\ref{60}) with  (\ref{47}) and (\ref{48}) we can express $D$ and $R$ as functions of $A$, $B$   
\begin{eqnarray}
\label{61a}
D({\bf k}) &=& \frac{1}{4} \frac{ B({\bf k}) - A({\bf k}) }{\sqrt{A({\bf k}) B({\bf k})}}  \\
\label{62}
R({\bf k}) &=& \frac{1}{2} \left\{ \frac{B({\bf k})+A({\bf k}) }{2 \sqrt{A({\bf k}) B({\bf k})}} -1 \right\}.
\end{eqnarray}
\noindent From (\ref{27}), (\ref{28}), (\ref{53}) and (\ref{54}) we have 
\begin{eqnarray}
\label{63}
A({\bf k}) - B({\bf k}) &=& 2 \int_{{\bf k}'} D_{{\bf k}'} \widetilde{V}({\bf k}' - {\bf k}) - \lambda \phi^{2} \\
\label{63A}
A({\bf k}) + B({\bf k}) &=& 2 (e(k) -\mu) + \lambda \phi^{2} + \int_{{\bf k}'} \left[\lambda + \widetilde{V}({\bf k}-{\bf k}') \right] R({\bf k}').
\end{eqnarray}
\noindent  For a given density $\rho$ of the system we can write  the constraint relation ${\cal N}/V = \rho$ which gives us
\begin{equation}
\label{63B}
\rho = \phi^{2} + \int_{{\bf k}'} R({\bf k}').
\end{equation}
To find a solution in the case $\phi=0$ we have to solve eqs. (\ref{63}), (\ref{63A}) and (\ref{63B}). Note that $R$ and $D$ are functions of $A$, $B$ through (\ref{61a}) and  (\ref{62}). With these three equations one can obtain $A$,$B$ and $\mu$.  
 
For the symmetry breaking solution $\phi \ne 0$, $\mu$ can be determined using the equation $B(0) = \lambda \phi^{2}$ which gives us
\begin{equation}
\label{mu}
\mu = {\cal U}_{{\rm e}}(0) + {\cal U}_{{\rm d}}(0) -{\cal U}_{{\rm p}}^{\rm r}(0) - \lambda \phi^{2}. 
\end{equation}
\noindent Which means that solving (\ref{63})-(\ref{mu}) gives us $A$, $B$ and $\phi$.   

\section{Equations of Motion in the Small Oscillation Regime Generalized RPA}

To find  the RPA equations we expand all the quantities around their equilibrium values, thus \footnote{In equations (\ref{64}) and (\ref{66}) the first delta is a Dirac delta and the second means an infinitesimal change.} 
\begin{eqnarray}
\label{64}
G({\bf k},{\bf k}',t)  &=&  G({\bf k})  \delta({\bf k}-{\bf k}')  + \delta G({\bf k},{\bf k}',t)  \\
\label{65}
\Sigma({\bf k},{\bf k}',t) &=& \delta{\Sigma}({\bf k},{\bf k}',t)  \\
\label{66}
\phi({\bf k},t) &=& \phi \delta({\bf k}) + \delta \phi({\bf k},t)  \\
\label{67}
\pi ({\bf k},t) &=& \delta \pi ({\bf k},t).
\end{eqnarray}
 \noindent For simplicity we write $G$ and $\Sigma$ in the basis where the equilibrium $G$ is diagonal and keep terms up to first order in small quantities. 
For the uniform case the diagonal basis will be plane waves. 
It will be useful to introduce new momentum coordinates so that
\begin{eqnarray}
\label{68}
{\bf P} &=& {\bf k} - {\bf k}' \\
\label{69}
{\bf q} &=& \frac{{\bf k}+{\bf k}'}{2} 
\end{eqnarray} 
\noindent so that 
 \begin{equation}
\label{70}
\delta G({\bf k},{\bf k}') \rightarrow \delta G({\bf P},{\bf q})
\end{equation}
\noindent and 
\begin{equation}
\delta G^{\ast}({\bf P},{\bf q}) = \delta G(-{\bf P},{\bf q}).
\end{equation}
\noindent Note that from (\ref{16}) the canonical variable  to $\delta G({\bf P},{\bf q},t)$ is $\delta \Sigma (-{\bf P},{\bf q})$. We will see that  ${\bf P}$ and ${\bf q}$ can be interpreted as total and relative momenta respectively so that we can write the RPA equations in the form where ${\bf P}$ is diagonal and can be considered as a dummy variable  
\begin{eqnarray}
\label{71}
\delta \dot{G}({\bf q},{\bf P},t) &=&\! \! \!s_{\rm K}({\bf q},{\bf P}) \delta \Sigma({\bf q},{\bf P},t) +  c_{\rm K} ({\bf q},{\bf P}) \delta \pi({\bf P},t) + \int_{{\bf q}'}S_{\rm K}({\bf q},{\bf q}',{\bf P}) \delta \Sigma({\bf q}',{\bf P},t)  \\
\label{72}
-\delta \dot{\Sigma}({\bf q},{\bf P},t) &=& \! \! \!s_{\rm M}({\bf q},{\bf P}) \delta G({\bf q},{\bf P},t) + c_{\rm M}({\bf q},{\bf P}) \delta \phi({\bf P},t) + \int_{{\bf q}'}S_{\rm M}({\bf q},{\bf q}',{\bf P}) \delta G({\bf q}',{\bf P},t) \\ 
\label{73}
\delta \dot{\phi}({\bf P},t) &=& \! \! \!   \delta \pi({\bf P},t) A({\bf P}) + \int_{{\bf q}'} c_{\rm K}({\bf q}',{\bf P}) \delta \Sigma({\bf q}',{\bf P},t) \\
\label{74}
-\delta \dot{\pi}({\bf P},t) &=& \! \! \! 
\delta \phi({\bf P},t) B({\bf P}) + \int_{{\bf q}'} c_{\rm M}({\bf q}',{\bf P}) \delta G({\bf q}',{\bf P},t). 
\end{eqnarray}
\noindent Where we note that the $(\pi,\phi)$ degrees of freedom are coupled to the much more numerous degrees of freedom $(\Sigma,G)$ which are labeled by ${\bf q}$. Different ${\bf q}$ values among $(\Sigma,G)$ are also coupled.
\noindent Introducing the notation
$f({\bf q}' + {\bf P}/2) = f'_{+}$ and $f({\bf q} - {\bf P}/2) = f_{-}$,  we find non-diagonal matrices in $({\bf q},{\bf q}')$ 
\begin{eqnarray}
\label{75}
S_{\rm K}({\bf q},{\bf q}',{\bf P})& = & 
2 \widetilde{V}({\bf q}-{\bf q}' ) \left[G_{+} G'_{+} + G_{-} G'_{-} \right] \\
\label{78}
S_{\rm{M}}({\bf q},{\bf q}',{\bf P}) & = &  \frac{\widetilde{V}({\bf q}-{\bf q}')}{2} +
\left[1 - \frac{G^{-1}_{+}
G^{-1}_{-}}{4} \right] \frac{\widetilde{V}({\bf P})}{4}
\left[1 - \frac{G^{'^{-1}}_{+}
G^{'^{-1}}_{1}}{4} \right] \nonumber \\
&& + \left[ \frac{G^{-1}_{+}
G^{-1}_{-}}{4} \right]
\frac{V({\bf q}-{\bf q}')}{2} \left[ \frac{G^{'^{-1}}_{+}G^{'^{-1}}_{-}}{4} 
\right]
\end{eqnarray}
\noindent and diagonal elements
\begin{eqnarray}
\label{76}
s_{\rm K}({\bf q},{\bf P})& = & 2 \left[ A_{+} 
G_{-} +  A_{-}
G_{+} \right]  \\ 
\label{79}
s_{\rm M}({\bf q},{\bf P})& = & \frac{G^{-2}_{+} G^{-1}_{-} A_{+} + G^{-2}_{-} G^{-
1}_{+}
A_{-}}{8}.
\end{eqnarray}
\noindent Finally we see the coupling elements between $(\pi,\phi)$ and $(\Sigma,G)$ 
\begin{eqnarray}
\label{77}
c_{\rm K}({\bf q},{\bf P})& = & \frac{\phi}{2} \left[G_{+} +
G_{-} \right] \left[ \widetilde{V}_{+} + 
\widetilde{V}_{-} \right]  \\
c_{\rm M}({\bf q},{\bf P})& = & \frac{\phi}{2} \left[ \tilde{V}_{+} + \tilde{V}_{-} + \tilde{V}({\bf P})\left(1 - \frac{G^{-1}_{+}
G^{-1}_{-}}{4} \right) \right], 
\end{eqnarray} 
\noindent which vanish when the symmetry is conserved $(\phi=0)$. As pointed out above the equations are diagonal in ${\bf P}$ so we can  interpret it as the total momentum of a pair of quasi-bosons, where $-{\bf k}'$ is thus interpreted as the momentum of a"hole". Because $\delta G$,  $\delta \Sigma$ and $\delta \phi$,  $\delta \pi$ are canonical variables we may invert the definitions of  momentum and coordinate, for reasons that will be clear in the next section, where the zero mode is discussed. We define column vectors
\begin{equation}
\label{81}
\Theta({\bf q},{\bf P},t) \equiv  \left( \begin{array}{c}
                                    \Theta_{\Sigma} \\
                                    \Theta_{\pi}
                                    \end{array} \right)
\equiv  \left( \begin{array}{c}
   \delta \Sigma({\bf q},{\bf P},t) \\
   \delta \pi({\bf P},t)
   \end{array}
                              \right), 
\; \; \; \; \Pi({\bf q},{\bf P},t) \equiv  \left( \begin{array}{c}
                                          \Pi_{G} \\
                                          \Pi_{\phi}  
                                          \end{array} \right)
                                    \equiv
                                        -\left( \begin{array}{c}
                                        \delta G({\bf q},{\bf P},t) \\
                                        \delta \phi({\bf P},t)
                                       \end{array}
                                \right).
\end{equation}
\noindent We can write a coupled oscillator Hamiltonian that corresponds to the RPA equations of motion in a matrix element form
\begin{equation}
\label{82}
H_{\rm RPA} = \frac{1}{2} \left( \Theta| {\cal A} |\Theta \right)   + \frac{1}{2} \left( \Pi| {\cal B}| \Pi \right),
\end{equation}
\noindent where the matrixes ${\cal A}$ and ${\cal B}$ are
\begin{equation}
\label{83}
{\cal A} =  \left( \begin{array}{cc}
              S_{\rm K} + s_{\rm K} & c_{\rm K} \\
              c_{\rm K}         & A
             \end{array}
     \right)
\; \; \; \; {\cal B}  = \left(\begin{array}{cc}
              S_{\rm M} + s_{\rm M} & c_{\rm M} \\
              c_{\rm M}         & B
                      \end{array}
      \right),
\end{equation}
\noindent and the equations of motion are
\begin{eqnarray}
\label{83a}
   | \dot{\Theta} ) &=&  {\cal B} |\Pi )  \\
\label{83b}
 |\dot{\Pi}) &=& - {\cal A}  |\Theta )
\end{eqnarray}
\nonumber which can be written as  second order equations
\begin{eqnarray}
\label{sec1}
| \ddot{\Theta} ) &=& - {\cal B} {\cal A} | \Theta ) \\
\label{sec2}
| \ddot{\Pi} ) &=& -{\cal A} {\cal B} |\Pi ).
\end{eqnarray}
We may separate the diagonal part of $H_{\rm RPA}$ so that
\begin{equation}
\label{84}
H_{\rm RPA} = H_{0} + H_{\rm int},
\end{equation}
\noindent where
\begin{equation}
\label{85}
H_{0} = \frac{1}{2}  \left(\matrix{\delta \Sigma^{\ast}&\delta \pi^{\ast}} \right)\left(\matrix{s_{\rm K}&0\cr 0&A} \right)\left(\matrix{\delta\Sigma\cr \delta \pi}\right)+\frac{1}{2} \left(\matrix{\delta G^{\ast}&\delta \phi^{\ast}} \right)\left(\matrix{s_{\rm M}&0\cr 0&B} \right)\left(\matrix{\delta G\cr \delta \phi}\right) 
\end{equation}
\noindent and $H_{\rm int}$ is equal to 
\begin{equation}
\label{85I}
H_{\rm int} = \frac{1}{2}  \left(\matrix{\delta \Sigma^{\ast}&\delta \pi^{\ast}} \right)\left(\matrix{S_{\rm K}&c_{\rm K}\cr c_{\rm K}&0} \right)\left(\matrix{\delta\Sigma\cr \delta \pi}\right)+\frac{1}{2} \left(\matrix{\delta G^{\ast}&\delta \phi^{\ast}} \right)\left(\matrix{S_{\rm M}&c_{\rm M}\cr c_{\rm M}&0}  \right)\left(\matrix{\delta G\cr \delta \phi}\right).
\end{equation}
\noindent Note that $A$ and $B$ must be positive for stability. If we introduce the trivial multiplicative canonical transformation
\begin{eqnarray}
\label{89}
\left(\matrix{\delta\Sigma\cr   \delta \pi}\right)  &\rightarrow & \left(\matrix{\delta\Sigma  \cr  \delta \pi}\right)' = 
\left(\matrix{\delta\Sigma \sqrt{s_{\rm M}}  \cr  \delta \pi \sqrt{B}}\right) \\
\label{90}
\left(\matrix{\delta G \cr \delta \phi} \right) &\rightarrow & \left(\matrix{\delta G  \cr  \delta \phi} \right)' =
\left(\matrix{\frac{\delta G}{\sqrt{s_{\rm M}}}  \cr   \frac{\delta \phi}{\sqrt{B}}} \right)
\end{eqnarray}
\noindent we obtain
\begin{equation}
\label{91}
H_{0} = \frac{1}{2}  \left(\matrix{\delta \Sigma &\delta \pi} \right)'\left(
\matrix{\Omega_{2}&0\cr 0&\omega} \right)\left(\matrix{\delta\Sigma\cr \delta \pi}\right)'+
\frac{1}{2} \left(\matrix{\delta G&\delta \phi} \right)'\left(\matrix{1&
0\cr 0&1} \right)\left(\matrix{\delta G\cr \delta \phi}\right)'.
\end{equation}
\noindent Where $\omega$ and $\Omega_{2}$ are 
\begin{eqnarray}
\label{93}
\omega({\bf P}) & =& \sqrt{A({\bf P}) B({\bf P})} \\
\label{94i}
\Omega_{2}({\bf q},{\bf P}) &=& \sqrt{s_{\rm K}({\bf q},{\bf P}) s_{\rm M}({\bf q},{\bf P})}. 
\end{eqnarray}
\noindent If we use the definitions of $s_{\rm K}$ and $s_{\rm M}$ from (\ref{76}) and ({\ref{79}) we get after same algebra the remarkable result
\begin{equation}
\label{94}
\Omega_{2}({\bf q},{\bf P}) = \sqrt{A_{+} B_{+}} + \sqrt{A_{-} B_{-}}= \omega({\bf k}) + \omega({\bf k}'),
\end{equation}
\noindent so that $\omega({\bf P})$ and $\Omega_{2}({\bf q},{\bf P})$ can be interpreted as 
the one and two free quasi-boson energies. We note that 
$\Omega_{2}(0,{\bf P}) = 2 \omega({\bf P}/2)$, which means that at zero relative momentum $\Omega_{2}({\bf P},0)$ corresponds to two quasi-bosons each  with momentum ${\bf P}/2$.
For the $\phi \ne 0$ case the free quasi-boson energies $\omega({\bf P})$ in general have a gap which means, 
that $\omega({\bf 0}) \ne 0$. In fact if we use (\ref{53}),(\ref{54}) and (\ref{mu}) we get
\begin{eqnarray}
A(0) &=& 2 {\cal U}_{p}^{\rm r}(0) + \lambda \phi^{2} \\
B(0) &=& \lambda \phi^{2}
\end{eqnarray}
\noindent because of (\ref{28}) we write
\begin{equation}
{\cal U}_{p}^{\rm r}(0) = \int_{{\bf k}'} \Delta({\bf k}') \widetilde{V}({\bf k}') = - \frac{\lambda}{2} \phi^{2} +
\int_{{\bf k}'} D({\bf k}') \widetilde{V}({\bf k}').
\end{equation}
\noindent We have then
\begin{equation}
\label{gap}
\omega(0) =  \sqrt{A(0) B(0)} = \sqrt{2 \lambda \phi^{2} \int_{{\bf k}'} V({\bf k}') D({\bf k}')}
\end{equation}

The oscillations of the $\delta \phi$, $\delta \pi$
pair can be interpreted as a quasi-boson mode with a gap while the oscillations
of $\delta G$, $\delta \Sigma$ can be interpreted as an interacting pair of
these same quasi-bosons. When $\phi=0$, we get $S_{3}= S_{6} = 0$ and the one
and two quasi-bosons systems are calculated independently. When $\phi \ne 0$
we must rediagonalize so that our final modes will be mixtures of one and
two quasi-bosons.
The variable $ { \bf q}$ represents the internal motion of the quasi-boson
pair
 with interaction given by the quantities $S$. In general this is a
scattering
problem and we must search for the scattering amplitude at a given energy  and
 ${\bf P}$, where the boundary conditions are determined by (\ref{94i}).
\noindent  We can write the equations for $\ddot{\delta G}$ and $\ddot{\delta \phi}$ using (\ref{71})-(\ref{74})  
\begin{eqnarray}
\label{RPA1}
\ddot{\delta G}({\bf q},{\bf P},t)&&+ \int_{{\bf q}',{\bf q}''} {\cal S}_{\rm K}({\bf q},{\bf q}',{\bf P}) {\cal S}_{\rm M}({\bf q}',{\bf q}'',{\bf P}) \delta G({\bf q}'',{\bf P},t) + c_{\rm K}({\bf q},{\bf P})  \int_{{\bf q}'} c_{\rm M}({\bf q}',{\bf P}) \delta G({\bf q}',{\bf P},t) \nonumber \\
&&= - \left[ B({\bf P}) c_{\rm K}({\bf q},{\bf P}) + \int_{{\bf q}'} {\cal S}_{\rm K}({\bf q},{\bf q}',{\bf P}) c_{\rm M}({\bf q}',{\bf P}) \right]  \delta \phi({\bf P},t) \\ 
\label{RPA2}
\ddot{\delta \phi}({\bf P},t)&&+ \left[ A({\bf P}) B({\bf P}) + \int_{{\bf q}'} c_{\rm K}({\bf q}',{\bf P}) c_{\rm M}({\bf q}',{\bf P}) \right] \delta \phi({\bf P},t) = \\
&&- A({\bf P}) \int_{{\bf q}'} c_{\rm M}({\bf q}',{\bf P}) \delta G({\bf q}',{\bf P},t) - \int_{{\bf q}',{\bf q}''} c_{\rm K}({\bf q}',{\bf P}) {\cal S}_{\rm M}({\bf q}',{\bf q}'',{\bf P})  \delta G({\bf q}'',{\bf P},t) \nonumber   
\end{eqnarray}
\noindent where ${\cal S}$ contain both $S$ and $s$ defined in (\ref{75})-(\ref{79}) i.e. 
\begin{eqnarray}
{\cal S}_{\rm K}({\bf q},{\bf q}',{\bf P})  &=& S_{\rm K}({\bf q}, ({\bf q}',{\bf P}) + s_{\rm K}({\bf q} ,{\bf q}',{\bf P}) \delta ({\bf q}'-{\bf q}) \\
{\cal S}_{\rm M}({\bf q},{\bf q}',{\bf P})  &=& S_{\rm M}({\bf q},
{\bf q}',{\bf P}) + s_{\rm M}({\bf q},{\bf q}',{\bf P}) \delta ({\bf q}'-{
\bf q}). 
\end{eqnarray}
\noindent If we look for  oscillatory solutions for the equations (\ref{RPA1}) and (\ref{RPA2}) 
\begin{eqnarray}
\delta G({\bf q},{\bf P},t) &=& \delta G({\bf q},{\bf P},t=0) e^{i \Omega t} \\
\delta \Sigma({\bf q},{\bf P},t) &=& \delta \Sigma({\bf q},{\bf P},t=0) e^{i \Omega t} \\
\delta \pi({\bf P},t) &=& \pi({\bf P},t=0) e^{i \Omega t} \\
\delta \phi({\bf P},t) &=& \phi({\bf P},t=0) e^{i \Omega t} 
\end{eqnarray}
\noindent we may   rewrite  (\ref{RPA1}) and (\ref{RPA2}) with the  compact mode notation of (\ref{sec2}) 
\begin{eqnarray}
\label{com1}
(\Omega^{2} - h_{2}) \Pi_{G} &=& \left(c_{\rm K} B + \int {\cal S}_{\rm K} \right) \Pi_{\phi} \\
\label{com2}
(\Omega^{2} - h_{1}) \Pi_{\phi} &=& \left( A \int c_{\rm M} + \int c_{\rm K} {\cal S}_{\rm M} \right) \Pi_{G}. 
\end{eqnarray}
\noindent  We have defined $h_{1}$ and $h_{2}$ as
\begin{eqnarray}
h_{1} &=& A B + \int_{\bf q} c_{\bf K} c_{\rm M} \\
h_{2} &=& {\cal S}_{\rm K} {\cal S}_{\rm M} + c_{\rm K} c_{\rm M} 
\end{eqnarray}
\noindent where $h_{1}$ depends only on ${\bf P}$ and $h_{2}$ is understood as a ${\bf P}$ dependent operator in ${\bf q}$ and ${\bf q}'$ with $c_{\rm K} c_{\rm M}$ being of separable form. We may  write $h_{1}$ and $h_{2}$ in terms of the free quasi-boson frequencies defined in (\ref{93}) and (\ref{94i}).
\begin{eqnarray}
h_{1} &=& \omega^{2} + \int_{\bf q} c_{\bf K} c_{\rm M} \\
h_{2} &=& \Omega_{2}^{2} + S_{\rm K} S_{\rm M} + s_{\rm K} S_{\rm M} + S_{\rm K} s_{\rm M} \equiv \Omega_{2}^{2} + V_{0}.  
\end{eqnarray}
\noindent We note that, in this form, $h_{2}$, (\ref{com1}) and (\ref{com2}) are not hermitian. However it can be shown that they correspond to a fully hermitian form which leads to a unitary $S$ matrix. We do not enter into this discussion here. \footnote{If this were a problem with a discrete spectrum the non hermitian structure has the form ${\cal A} {\cal B}$ with ${\cal A}$ and ${\cal B}$ both hermitian but non commuting. It is well known that the secular determinant Det$|\Omega^{2} - {\cal A}{\cal B}|$ is equivalent to that of the hermitian form Det$|\Omega^{2} -\sqrt{{\cal A}} {\cal B} \sqrt{\cal A}|$, corresponding to a transformation of the variables $\Pi$.} 

Eliminating $\Pi_{\phi}$ from (\ref{com1}) and (\ref{com2}) we get 
\begin{equation}
\left[ \left( \Omega^{2} - h_{2} \right) - \left(c_{\rm k} B + {\cal S}_{\rm K} c_{\rm M} \right) \frac{1}{(\Omega^{2} - h_{1}+ i \epsilon)} \left(A \int c_{\rm M} +  \int c_{\rm K} {\cal S}_{\rm M} \right) \right] \delta G = 0 
\end{equation}
\noindent Defining the separable potential
\begin{equation}
V_{1} \equiv \left(c_{\rm k} B + {\cal S}_{\rm K}c_{\rm M} \right) \frac{1}{(\Omega^{2} - h_{1}+i \epsilon)} \left(A  c_{\rm M} +  c_{\rm K} {\cal S}_{\rm M} \right)\end{equation}
\noindent we can write 
\begin{equation}
\left(\Omega^{2} - \Omega_{2}^{2} \right) = \left( V_{0} + V_{1} \right) \delta G
\end{equation}
\noindent which lead us to the standard form for the scattering for a given  value for $\Omega= \Omega_{2}({\bf P},{\bf q})$ and $\Omega_{2}'= \Omega_{2}({\bf P},{\bf q}')$.
\begin{equation}
\label{102}
\Pi_{G}({\bf q},{\bf q}') = \delta({\bf q}-{\bf q}') +   \frac{1}{\Omega^{2} - \Omega_{2}'^{2} - i \epsilon}   V \Pi_{G} \equiv  \delta({\bf q}-{\bf q}') + {\cal G}_{0}^{(+)} V \Pi_{G}.
\end{equation}
\noindent In the usual fashion we can define a $T$ matrix as
\begin{equation}
T  = V \Pi_{G} 
\end{equation}
\noindent so that the equation that determines the matrix $T$ is
\begin{equation}
\label{fscattering}
T = V + V {\cal G}_{0}^{(+)} T. 
\end{equation}
\noindent In principle for a given interaction we can solve for the $T$ matrix obtaining all the scattering 
properties like phase shifts and bound states. 
The existence of bound states will be determined by the poles of the $T$ matrix. Because of the coupling between one and two quasi-bosons there can  be one "bound" state which 
 can be interpreted as the  dispersion relation of a new RPA-boson $\Omega_{\rm b}$. Again using (\ref{com1}) and (\ref{com2}) we can eliminate $\delta G$ in favor of $\delta \phi$ and obtain a formal form for the dispersion relation for $\Omega_{\rm b}({\bf P})$ 
\begin{equation}
 \left( \Omega_{\rm b}^{2} - h_{1} \right) - \left( A c_{\rm M}  + c_{\rm K} {\cal S}_{\rm M}
 \right) \frac{1}{\Omega^{2} - h_{2}} \left(B  c_{\rm K} +  c_{
\rm M} {\cal S}_{\rm K} \right)   = 0.
\end{equation}
\noindent This involves the Greens function for $h_{2}$ below the threshold so that the $i \epsilon$ in not necessary. Because of the broken symmetry (see next section) 
this dispersion relation will start at zero, in contrast to $\omega({\bf P})$ given in (\ref{93}). This can be seen 
schematically in Fig. 1. If the interaction is attractive enough there can of course be additional bound states. If any bound state actually leads to a negative value for $\Omega^{2}$ the system is unstable, as is usual for the RPA.
\section{The Goldstone mode}
In order to understand the structure of the Goldstone mode we recognize from (\ref{sec1}) that for $\Omega=0$ we must have 
\begin{equation}
\label{g2}
{\cal A} |\theta^{(0)} )   = 0,
\end{equation}
\noindent going back to the more explicit notation we have
\begin{eqnarray}
\label{g3}
&& \int_{{\bf q}'}S_{\rm K}({\bf q},{\bf q}',{\bf P})  \theta_{\Sigma}^{(0)}({\bf q}',{
\bf P}) + s_{\rm K}({\bf q},{\bf P}) \theta_{\Sigma}^{(0)} ({\bf q},{\bf P}) + c_{\rm K}({\bf q},{\bf P}) \theta_{\pi}^{(0)}({\bf
P}) = 0  \\
\label{g4}
&& \int_{{\bf q}'} c_{\rm K} ({\bf q}',{\bf P}) \theta_{\Sigma}^{(0)}({\bf q}',{\bf P}) +
\theta_{\pi}^{(0)}({\bf P}) A({\bf P}) = 0. 
\end{eqnarray}
\noindent Eliminating $\theta_{\pi}^{(0)}({\bf P})$ from (\ref{g4}) we get an equation for $\theta_{\Sigma}^{(0)} $ 
\begin{equation}
\label{g5}
\int_{{\bf q}'}S_{\rm K}({\bf q},{\bf q}',{\bf P}) \theta_{\Sigma}^{(0)}({\bf q}',{
\bf P}) + s_{\rm K} ({\bf q},{\bf P}) \theta_{\Sigma}^{(0)} ({\bf q},{\bf P}) -\frac{1}{A({\bf P})} c_{\rm K}({\bf q},{\bf P}) \int_{{\bf q}'} c_{\rm K}({\bf q}',{\bf P}) \theta_{\Sigma}^{(0)}({\bf q}',{\bf P}) =0.
\end{equation}
\noindent If ${\bf P} =0$ we can find a solution for (\ref{g5}) 
\begin{equation}
\label{g6}
\theta_{\Sigma}^{(0)}({\bf q},0) = 1 - \frac{G^{-2}({\bf q},0)}{4},
\end{equation}
\noindent which arises from the properties of the static solutions. To do so we substitute (\ref{g6}) in ({\ref{g5}) and  using 
the static properties (\ref{63})-(\ref{63A}) we compute each one of the integrals separately. For convenience we omit the index $({\bf P}=0)$ 
\begin{eqnarray}
\label{g7}&&\int_{{\bf q}'}S_{\rm K}({\bf q},{\bf q}')\left[1 - \frac{G^{-2}({\bf q})}{4} \right]\;\;\; = \;\;\;4 G({\bf q}) \int_{{\bf q}'} \widetilde{V}({\bf q}-{\bf q}') \left[G({\bf q}')- \frac{G^{-1}({\bf q}')}{4} \right] \nonumber \\ 
&&= 4 G({\bf q}) \left[B({\bf q}) -A({\bf q}) -\phi^{2} \widetilde{V}({\bf q})\right]
\end{eqnarray}
\noindent and
\begin{eqnarray}
\label{g8}
&&\int_{{\bf q}'} c_{\rm K}({\bf q}')\left[1 - \frac{G^{-2}({\bf q})}{4} \right]\;\;\; =\;\;\; 2 \phi \int_{{\bf q}'} \widetilde{V}({\bf q}')\left[G({\bf q}')- \frac{G^{-1}({\bf q}')}{4}\right] \nonumber \\
&&= 2 \phi \left[B(0) - A(0) - \lambda \phi^{2}  \right] =  -2 \phi A(0).
\end{eqnarray}
\noindent It is easy to see that the subtitution of (\ref{g7}) and (\ref{g8})  leads to satisfying (\ref{g5}). Once we know $\theta_{\Sigma}^{(0)}$ we can find $\theta_{\pi}^{(0)}$ using (\ref{g4}) so that
\begin{equation}
\theta_{\pi}^{(0)} = - \frac{1}{A(0)} \int_{{\bf q}'} c_{\rm K} ({\bf q}')\left[1 - \frac{G^{-2}({\bf q})}{4} \right] = 2 \phi.
\end{equation}
\noindent This gives us a particular normalized  column vector  $\theta^{(0)}$ for the zero mode
\begin{equation}
\theta^{(0)} =  C \left(\matrix{
              \frac{1}{2} \left[1 - \frac{G^{-2}({\bf q},0)}{4} \right] \cr
              \phi} 
              \right).
\end{equation}
\noindent with  
\begin{equation}
C = \frac{1}{\sqrt{\int_{{\bf q}'}\frac{1}{2} \left[1 - \frac{G^{-2}({\bf q},0)}{4} \right]^{2} + \phi^{2}}}.
\end{equation}
\noindent This corresponds to a new coordinate and momentum which are linear combinations of the previous $(\delta G, \delta \phi)$ and $(\delta \Sigma, \delta \pi)$ and can be written as inner products.
\begin{eqnarray}
{\cal Q} &\equiv& ( \theta^{(0)} | \Theta ) = C \left(\matrix{\frac{1}{2} \left[1 - \frac{G^{-2}({\bf q},0)}{4} \right], & \phi } \right)\left(\matrix{\delta\Sigma\cr \delta \pi}\right), \\
{\cal P} &\equiv& ( \theta^{(0)} | \Pi ) =    C \left(\matrix{\frac{1}{2} \left[1 - \frac{G^{-2}({\bf q},0)}{4} \right], & \phi }\right)\left(\matrix{\delta G\cr \delta \phi}\right). 
\end{eqnarray}
\noindent To better understand these coordinates we expand ${\cal N}$, the mean number of particles, up to first order
\begin{equation}
{\cal N} = {\cal N}_{0} + \delta {\cal N},
\end{equation}
\noindent where ${\cal N}_{0}$ is the static result given by
\begin{equation}
{\cal N}_{0} = \frac{1}{2} \int_{{\bf q}'} \left[ \frac{G^{-1}({\bf q}',0)}{4} + G({\bf q}',0) -1 \right] + \frac{\phi^{2}}{2}.
\end{equation}
\noindent The first order term $\delta {\cal N}$ is
\begin{equation}
\delta {\cal N} = \frac{1}{2} \int_{{\bf q}'} \left(1 - \frac{G^{-2}({\bf q}',0)}{4
} \right) \delta G({\bf q}',0) + \phi \delta \phi.
\end{equation}
\noindent Which we see is proportional to our new canonical momentum 
\begin{equation}
\label{mo}
{\cal P} = C \delta {\cal N}.
\end{equation}
\noindent From (\ref{83b}) and (\ref{g2}) we have 
\begin{equation}
(  \theta^{(0)} \dot{| \Pi )} = \dot{{\cal P}} = - ( \theta^{(0)} |{\cal A} | \Theta ) = 0.
\end{equation}
\noindent Indeed this is the general result $(\dot{\delta {\cal N}} =0)$ proved in section 1 and, as expected, the zero mode corresponds to a "translational" motion in N with no oscillation. 

In fact because we have the RPA Hamiltonian (\ref{82}) it is possible to see the explicit form for the mass 
coefficient ${\cal M}$ in the quadratic expression for the $\delta {\cal N}$ dependence. To see this  we first 
rewrite the RPA Hamiltonian (\ref{82}) in a basis where ${\cal A}$ is diagonal, because it contains  the zero mode.
\begin{equation}
H_{\rm RPA} = \frac{1}{2} \theta^{i^\ast} ( i | {\cal A} |i ) \theta^{i} + \frac{1}{2} P^{i^\ast} ( i |{\cal B}|j ) P^{j}.
\end{equation}
\noindent with
\begin{equation}
( 0 | {\cal A}| 0 ) = 0.
\end{equation}
\noindent  This shows that there is no dependence on $\theta^{(0)}$ which corresponds to our original parameter $\theta$ canonical to ${\cal N}$. The summation on $(i,j)$ is implicit. We now separate the contributions $P^{(0)}$ from  the kinetic piece as
\begin{eqnarray}
\label{P0}
{\cal K} \equiv \frac{1}{2} P^{i^\ast} (  i |{\cal B}|j )  P^{j} &=&\frac{1}{2} P^{m^\ast} ( m|{\cal B}|m' ) P^{m'} +  \frac{1}{2} P^{0^\ast} ( 0 |{\cal B}|m )  P^{m} \nonumber \\
&&+ \frac{1}{2} P^{m^\ast} (  m |{\cal B}|0 )  P^{0}
 + \frac{1}{2} P^{0^\ast} ( 0 |{\cal B}|0 )  P^{0}   
\end{eqnarray}
\noindent where $m$ and $m'$ are different from zero. We can diagonalize ${\cal B}$ in the subspace getting 
\begin{eqnarray}
{\cal K} &=& \frac{1}{2}
P^{\lambda^\ast} ( \lambda |{\cal B}|\lambda ) P^{\lambda} +  \frac{1}{2} P^{0^\ast} ( 0 |{\cal B}|\lambda )
 P^{\lambda} \nonumber 
+ \frac{1}{2} P^{\lambda^\ast} (  \lambda |{\cal B}|0 )  P^{0} \\
&& + \frac{1}{2} P^{0^\ast} ( 0 |{\cal B}|0 )  P^{0}.
\end{eqnarray}
Using the fact that  $P^{0}$ is constant  we  complete squares 
\begin{eqnarray}
{\cal K} &=&\frac{1}{2} \left[ P^{\lambda} + P^{0} X^{\lambda} \right]^{\ast}
 ( \lambda |{\cal B}| \lambda ) \left[ P^{\lambda} +  X^{\lambda} P^{0} \right]+
\frac{1}{2} P^{0^\ast} ( 0|{\cal B}|0 ) P^{0} \nonumber \\
&&- \frac{1}{2} P^{0^\ast} X^{\lambda^\ast} ( \lambda |{\cal B}| \lambda ) X^{\lambda^\ast} P^{0}
\end{eqnarray}
\noindent where the vector  $X$ has to be equal to
\begin{equation}
X^{\lambda} = ( 0|{\cal B}|\lambda ) \frac{1}{( \lambda | {\cal B} | \lambda )}.
\end{equation}
\noindent So that the coefficient of $(P^{0})^{2}/2$ will be
\begin{equation}
\frac{1}{\cal M} = ( 0|{\cal B}|0 ) - ( 0 |{\cal B}|\lambda )  \frac{1}{(\lambda |{\cal B}|\lambda )} (\lambda |{\cal B}| 0) 
\end{equation}
\noindent which gives us
\begin{equation}
\label{Mass}
{\cal M} = \frac{1}{( 0|{\cal B}|0 ) - ( 0 |{\cal B}|\lambda )  \frac{1}{( \lambda |{\cal B}| \lambda)} ( \lambda|{\cal B}| 0 )} \equiv  ( 0 | \frac{1}{\cal B} | 0 ) \ne \frac{1}{( 0 | {\cal B} |0 )}.
\end{equation}
\noindent Note that this a non-trivial result, and it is the effect of  the linear terms in $P^{0}$ in (\ref{P0}). If we use (\ref{mo}) we can actually see that
\begin{equation}
\label{deltaN1}
\frac{(P^{0})^{2}}{2 {\cal M}} = \frac{C^{2} {\delta {\cal N}}^{2}}{2 {\cal M}} 
\end{equation}
\noindent which means we have calculated the coefficient of ${\delta {\cal N}}^{2}$. To see the physical meaning of this coefficient we expand our original ${\cal H} - \mu {\cal N}$ around the equilibrium value ${\cal N}_{0}$
\begin{equation}
{\cal H}({\cal N}) - \mu({\cal N}_{0}) {\cal N} = {\cal H}({\cal N}_{0}) + \delta {\cal N} \frac{d {\cal H}}{d {\cal N}_{0}} + \frac{1}{2} {\delta {\cal N}}^{2} \frac{d^{2} {\cal H}}{d {\cal N}_{0}^{2}}  - \mu({\cal N}_{0}) {\cal N}+ \cdots .
\end{equation}
\noindent Using  
\begin{equation}
\mu = \frac{d {\cal H}}{d {\cal N}_{0}}
\end{equation}
\noindent we have
\begin{equation}
\label{deltaN2}
{\cal H}({\cal N}) - \mu({\cal N}_{0}) {\cal N} = {\cal H}({\cal N}_{0}) - \mu({\cal N}_{0}) {\cal N}_{0} + 
\frac{1}{2} {\delta {\cal N}}^{2} \frac{d^{2} {\cal H}}{d {\cal N}_{0}^{2}}
\end{equation}
\noindent comparing the ${\delta {\cal N}}^{2}$ coefficients in (\ref{deltaN1}) and (\ref{deltaN2}) we have
\begin{equation}
\frac{d^{2} {\cal H}}{d {\cal N}_{0}^{2}} = \frac{d \mu}{d {\cal N}_{0}} =  \frac{C^{2}}{\cal M}   
\end{equation}
\noindent Which means that once we have computed ${\cal M}$ using (\ref{Mass}) we  also get  the value of
 the second derivative of the energy with  respect to the number of particles.

For the remaining modes we have to work with a Hamiltonian in the sub-space that excludes the zero mode i.e. 
\begin{equation}
H_{\rm RPA}^{\rm s} = \frac{1}{2} P^{i^\ast}_{\rm s} \left(i| {\cal B}_{\rm s}|j \right) P^{j}_{\rm s} + \frac{1}{2} Q^{i^\ast}_{\rm s} \left( i |{\cal A}_{\rm s}| j \right) Q^{j}_{\rm s}.
\end{equation}
\noindent If we introduce the canonical transformation
\begin{eqnarray}
P_{\rm s} &\rightarrow & \frac{1}{\sqrt{{\cal B}_{\rm s}}}  P_{\rm s}   \\
Q_{\rm s} &\rightarrow & \sqrt{{\cal B}_{\rm s}} Q_{\rm s}
\end{eqnarray}
\noindent we get
\begin{equation}
H_{\rm RPA}^{\rm s} = \frac{1}{2} P^{i^\ast}_{\rm s} P^{i}_{\rm s} + \frac{1}{2} Q^{i^\ast}_{\rm s} \left(i | \sqrt{{\cal B}_{\rm s}} {\cal A}_{\rm s} \sqrt{{\cal B}_{\rm s}} |j \right) Q^{j}. 
\end{equation}
\noindent If we now diagonalize the matrix $ \sqrt{{\cal B}_{\rm s}}  {\cal A}_{\rm s} \sqrt{{\cal B}_{\rm s}} $ we have the final form to be used in subsection 6.2.
\begin{equation}
\label{finalRPA}
H_{\rm RPA}^{\rm s} = \frac{1}{2} {\cal P}^{\Omega^\ast} {\cal P}^{\Omega} + \frac{1}{2} {\cal Q}^{\Omega^\ast} \Omega^{2} {\cal Q}^{\Omega},
\end{equation}
\noindent where $\Omega^{2}$ is real for stability and the ${\cal P}^{\Omega}$ and ${\cal Q}^{\Omega}$ are in general complex.

\section{General Remarks}
\subsection{Connection with the Bogoliubov Transformation}
One can construct the operator that  annihilates  our trial wave functional (\ref{9}) so that  
\begin{equation}
\hat{\xi}(t) \Psi[\phi',t] =0
\end{equation}
\noindent where
\begin{equation}
\label{B1}
\hat{\xi}(t) = \frac{1}{2}\left[\frac{1}{G^{1/2}} - 4 i G^{1/2} \Sigma \right]  \delta \hat{\phi} + i \sqrt{2} G^{1/2} 
 \delta \hat{\pi} 
\end{equation}
\noindent and  
\begin{equation} 
\delta \hat{\phi} = \hat{\phi} -\phi, \;\;\; \delta \hat{\pi}  = \hat{\pi} -\pi. 
\end{equation}
\noindent  Note that this result is valid for any particular time $t$. For the present purpose we will use the static annihilation operator   
\begin{equation}
\hat{\xi}(0) \equiv \hat{\xi} = \frac{1}{2} \frac{1}{G^{1/2}} \delta \hat{\phi} + i G^{1/2} \hat{\pi} 
\end{equation}
\noindent where now both $G$ and $\phi$ are the time independent equilibrium results. This will be our quasi-boson annihilation operator. The corresponding creation operator is then  
\begin{equation}
\hat{\xi}^{\dag} = \frac{1}{2} \frac{1}{G^{1/2}} \delta \hat{\phi}  -  i G^{1/2} \hat{\pi} 
\end{equation}
\noindent and as usual 
\begin{eqnarray}
\label{R0}
&& [\hat{\xi}_{\rm a},\hat{\xi}_{\rm b}^{\dag}] = \delta_{{\rm a},{\rm b}} \\
&& [\hat{\xi}_{\rm a},\hat{\xi}_{\rm b}] = [\hat{\xi}_{\rm a}^{\dag},\hat{\xi}_{\rm b}^{\dag}] = 0
\end{eqnarray}
\noindent then
\begin{equation}
\label{R1}
\delta \hat{\phi}= \sqrt{G} \left[\hat{\xi} + \hat{\xi}^{\dag} \right] 
\end{equation}
\noindent and 
\begin{equation}
\label{R11}
\hat{\pi} = - \frac{i}{2 \sqrt{G}} \left[\hat{\xi} - \hat{\xi}^{\dag} \right].
\end{equation}
We can now make the connection with the Bogoliubov transformation 
writing  the quasi-boson operator  as a function of the original $\hat{\psi}$ as
\begin{equation}
\label{R111}
\hat{\xi} = X \left[\hat{\psi} - \psi \right] + Y \left[\hat{\psi}^{\dag} - \psi^{\ast} \right]
\end{equation}
\noindent with
\begin{eqnarray}
X &=& \frac{1}{2 \sqrt{2}} \left[ \frac{1}{G^{1/2}} + 2 G^{1/2} \right]  \\
Y &=& \frac{1}{2 \sqrt{2}} \left[ \frac{1}{G^{1/2}} -2 G^{1/2} \right] 
\end{eqnarray}
\noindent Equation (\ref{R111}) is a generalized version of the Bogoliubov transformation where the 
commutation rules (\ref{R0}) imply that the matrix relation
\begin{equation}
X^{2}  -  Y^{2}  = 1
\end{equation}

as
\begin{eqnarray}
D &=& X Y \\
R &=& Y^{2} 
\end{eqnarray}

We may write the original Hamiltonian in terms of these  quasi-boson  operators $\hat{\xi}$ and $\hat{\xi}^{\dag}$. It is easy to
see that there will be  terms like $\phi \hat{\xi} \hat{\xi} \hat{\xi}^{\dag}$ which turn two quasi-bosons into one justifying our interpretation of the generalized RPA. In the next subsection we will show how the small oscillations correspond to modes that are mixtures of one and two quasi-bosons.
\subsection{One and two quasi-boson coupling}
\noindent  We proceed by expanding our wave functional  around the equilibrium values for $G$, $\Sigma$, $\phi$ and $\pi$ 
\begin{equation}
\langle \phi'|\Psi(t)| \rangle \equiv \Psi[\phi',t] \approx  \left[ 1 - \int \delta \phi' \left(\frac{1}{4} \frac{1}{G} \delta G \frac{1}{G} -
i \delta \Sigma \right) \delta  \phi' - \int \delta \phi' \left(\frac{1}{2 G} \delta \phi - i \delta \pi \right) \right] \Psi_{0}[\phi']
\end{equation}
\noindent where $\Psi_{0}[\phi']$ is the ground state wave functional that is annihilated by $\hat{\xi}$. If we use (\ref{R1}) and (\ref{R11}) we can write
\begin{eqnarray}
\Psi[\phi',t] &\approx&  \left\{ 1 -\int \left( \frac{1}{4} \frac{1}{\sqrt{G}} \delta G \frac{1}{\sqrt{G}} - i \sqrt{G} \delta \Sigma \sqrt{G} \right) \left(  \hat{\xi}^{\dag} \hat{\xi}^{\dag} + \hat{\xi} \hat{\xi} + \hat{\xi} \hat{\xi}^{\dag} + \hat{\xi}^{\dag} \hat{\xi} \right) \right. \nonumber \\ 
&&\left. - \int \left( \frac{1}{2 \sqrt{G}} \delta \phi - i \sqrt{G} \delta \pi \right) \left(\xi^{\dag} + 
\xi \right)  
\right\} \Psi_{0}[\phi']. 
\end{eqnarray}
\noindent Note that the equilibrium $G$ can be taken diagonal in a particular representation while in this same representation $\delta G$ and $\delta \Sigma$ are non-diagonal. If we expand $\delta G$ , $\delta \phi$, $\delta \Sigma$ and $\delta \pi$ in the $\Omega$ modes found by solving the RPA equations we have the general structure. Thus 
\begin{eqnarray}
\left( \matrix{\delta \Sigma \cr \delta \pi} \right) &=& \sum_{\Omega} {\cal P}^{\Omega}
\left( \matrix{ \theta_{\Sigma}^{\Omega} \cr \theta_{\pi}^{\Omega} } \right) e^{i \Omega t} \\ 
\left( \matrix{\delta G \cr \delta \phi} \right) &=& \sum_{\Omega} {\cal Q}^{\Omega} \left( \matrix{ \theta_{\Sigma}^{\Omega} \cr \theta_{\pi}^{\Omega}} \right) e^{i \Omega t}
\end{eqnarray}
\noindent with the hamiltonian (\ref{finalRPA}) for ${\cal P}^{\Omega}$ and ${\cal Q}^{\Omega}$. The state vector looks like
\begin{equation}
| \Psi(t) \rangle \approx \left\{ |0 \rangle  + \sum_{\Omega} \left[  |1\rangle +  |2 \rangle \right] e^{i \Omega t } \right\}
\end{equation}
\noindent where we have taken into account that $\hat{\xi}$ annihilates the ground state and have defined
\begin{eqnarray}
|1 \rangle &=& \int \Phi^{\Omega}_{1} \xi^{\dag} |0 \rangle \\
|2 \rangle &=& \int \Phi^{\Omega}_{2} \xi^{\dag} \xi^{\dag}|0 \rangle
\end{eqnarray}
\noindent where $\Phi^{\Omega}_{1}$ and $\Phi^{\Omega}_{2}$ are the  one and two particle wave functions given by
\begin{eqnarray}
\Phi^{\Omega}_{1} &=& {\cal P}^{\Omega} \frac{1}{2 \sqrt{G}} \theta_{\pi}^{\Omega}- i {\cal Q}^{\Omega} \sqrt{G} \theta_{\pi}^{\Omega} \\
\Phi^{\Omega}_{2} &=& {\cal P}^{\Omega}  \frac{1}{2 \sqrt{G}} \theta_{\Sigma}^{\Omega} \frac{1}{2 \sqrt{G}} -i {\cal Q}^{\Omega} 
 \sqrt{G} \theta_{\Sigma}^{\Omega}  \sqrt{G}.
\end{eqnarray}
 Note that for each value of $\Omega$ (either the bound state $\omega$ or the the continuum $\Omega_{2}$ ) we have, as expected, a mixture of one and two quasi-bosons.   
We take into account that for each given mode \footnote{In fact as we note before in section 5 this relation does not hold for the zero mode where ${\cal P}^{0}$ = constant and ${\cal Q}^{0} = ({\cal P}/{\cal M}) t$} have a relationship between ${\cal P}^{\Omega}$ and ${\cal Q}^{\Omega}$
\begin{equation}
{\cal Q}^{\Omega} = \frac{{\cal P}^{\Omega}}{i \Omega}.
\end{equation}
\noindent Then the unnormalized wave functions can be written as
\begin{eqnarray}
\Phi^{\Omega}_{1} &=&  \frac{1}{2 \sqrt{G}}  \theta_{\pi}^{\Omega} - \frac{\sqrt{G}}{\Omega} \theta_{\pi}^{\Omega} \\
\Phi^{\Omega}_{2} &=& \frac{1}{2 \sqrt{G}} \theta_{\Sigma}^{\Omega} \frac{1}{2 \sqrt{G}} -
\frac{1}{\Omega} \sqrt{G} \theta_{\Sigma}^{\Omega}  \sqrt{G}. 
\end{eqnarray} 
In order to understand the significance of the bound state for ${\bf P}=0$ we can examine its time independent  two body component, $\Phi^{\omega}({\bf P}=0,{\bf q})$. In this particular case   
\begin{equation}
\Phi^{0}_{2} = \frac{1}{\sqrt{G}} \theta_{\Sigma}^{0} \frac{1}{\sqrt{G}}.
\end{equation}
\noindent From (\ref{g6}) we get
\begin{equation}
\Phi^{0}_{2} = \frac{1}{\sqrt{G({\bf q})}}  \left(1 - \frac{1}{4 G^{2}({\bf q})} \right)  \frac{1}{\sqrt{G({\bf q})}}.   
\end{equation}
\noindent If it is to  represent the  bound state of a pair of quasi-bosons  the integral 
\begin{equation}
\label{conv}
\int_{{\bf q}} (\Phi^{0}_{2})^{2}    = \int_{{\bf q}} \frac{B({\bf q})}{A({\bf q})} \left[ \frac{A({\bf q})-B({\bf q})}{A({\bf q})^{2}} \right] 
\end{equation}
\noindent must be finite. To show this, we examine the behaviour of $(\Phi^{0}_{2})^{2}$. For ${\bf q} \rightarrow 0$, it  goes to a finite value because of the gap as discussed in (\ref{gap}). For ${\bf q} \rightarrow \infty$ all the generalized potentials go to zero and both $A$ and $B$ go to infinity as $q^{2}$ plus a constant. This leads to   $(\Phi^{0}_{2})^{2}  \rightarrow \alpha/k^{4}$ where $\alpha$ is the difference of these constants. This behavior assures the convergence of  (\ref{conv}).    

For the case when ${\bf  P} \ne 0$ we expect this normalizable bound state to evolve with ${\bf P}$ and to exist for at least a finite range of ${\bf  P}$. We note that it is not separable in ${\bf P}$ and ${\bf  q}$ so that the relative wave function changes with ${\bf  P}$ and could become unbound for some critical ${\bf P}$.
\section{Temperature Dependent Calculation}
Temperature is often  introduced \cite{TT} in the so called temperature dependent Hartree Fock Bogoliubov approximation by generalizing  the static
quantities $R$ and $D$ defined in (\ref{47}) and (\ref{48}) as 
\begin{eqnarray}
D({\bf  P}) &=& X({\bf  P}) Y({\bf  P}) \rightarrow  X({\bf  P}) Y({\bf  P}) (1 + 2 \nu({\bf  P})) \\ 
R({\bf  P}) &=& Y^{2}({\bf  P}) \rightarrow X^{2}({\bf  P}) \nu({\bf  P}) + Y^{2}({\bf  P}) (1 + \nu({\bf   P})) 
\end{eqnarray}
\noindent with  the occupation $\nu({\bf  P })$  given by 
\begin{equation}
\nu({\bf P}) = \frac{1}{e^{\omega(P)/KT}-1}
\end{equation}
\noindent This corresponds  to the usual grand canonical ensemble where $\omega$ is the one  quasi-boson energy defined in (\ref{93}), with its gap. 
This obviously describes the physics as having free quasi-bosons where neither the existence of the gapless bound state nor the interaction between the quasi-bosons is  taken into account. 

Our approach to the problem is different in that we use the canonical ensemble. We have seen that the RPA Hamiltonian consists of a free quasi-boson part $H_{0}$ (\ref{85}) and $H_{\rm int}$ (\ref{85I}) that treats the interaction between the quasi-bosons. So our first approximation in the canonical ensemble  for the free energy will be
\begin{equation}
\label{f1}
F^{(1)}({\cal N},T) =  {\cal H}  +  \frac{1}{\beta} \int_{\bf P}\log(1 - e^{- \beta \omega({\bf P})}).
\end{equation}
\noindent where  $\omega$ is the free quasi-boson energy given in (\ref{93}). We can also extract more information due to the interaction of the quasi-bosons and this can be done by computing the second virial coefficient \cite{RD} 
\begin{equation}
\label{f2}
F^{(2)}({\cal N},T) = \frac{1}{\beta}  \int_{\bf P}  e^{- \beta \Omega({\bf P})} +  \frac{1}{4 \pi i} \int_{{\bf P},{\bf q}}   e^{- \beta \Omega_{2}} Tr S^{-1} \frac{\stackrel{\leftrightarrow}{\partial}}{\partial \Omega_{2}} S. 
\end{equation}
\noindent Where $\Omega({\bf P})$ is the zero mode bound state dispersion relation and $S$ can be determined, as mention before, through the scattering of the quasi-bosons using (\ref{fscattering}).
Note that for the cases where we have two variational solutions  $\phi =0$ and  $\phi \ne 0$ we can calculate using (\ref{f1}) and (\ref{f2}) for a given ${\cal N}$ obtaining  two free energies and this allow us to determine the critical temperature by setting their derivatives with respect to ${\cal N}$ equal i.e looking for the same value of the chemical potential \cite{SM}

{\large{Figure Caption}

Fig. 1: Schematic plot of the one $\omega$ and two $\Omega_{2}$ free quasi-boson as a function of the total momentum. When the interaction between them is taken in account (RPA) we have a scattering region and a gapless bound state $\Omega$
\end{document}